# Anomalous Polarization Reversal in Strained Thin Films of CuInP$_2$S$_6$

Anna N. Morozovska[1*], Eugene A. Eliseev[2], Ayana Ghosh[3], Mykola E. Yelisieiev[4], Yulian M. Vysochanskii[5†], and Sergei V. Kalinin[6‡]

[1] Institute of Physics, National Academy of Sciences of Ukraine,

Nauky Avenue, Bldg. 46, 03028 Kyiv, Ukraine

[2] Institute for Problems of Materials Science, National Academy of Sciences of Ukraine,

Krjijanovskogo Street, Bldg. 3, 03142 Kyiv, Ukraine

[3] Computational Sciences and Engineering Division, Oak Ridge National Laboratory, Oak Ridge,

Tennessee, 37831

[4] Physics Faculty of Taras Shevchenko National University of Kyiv, Volodymyrska street 64, Kyiv,

01601, Ukraine

[5] Institute of Solid-State Physics and Chemistry, Uzhhorod University,

88000 Uzhhorod, Ukraine

[6] Department of Materials Science and Engineering, University of Tennessee,

Knoxville, TN, 37996, USA

**Abstract**

Strain-induced transitions of polarization reversal in thin films of a ferrielectric CuInP$_2$S$_6$ with ideally-conductive electrodes is explored using the Landau-Ginzburg-Devonshire approach with an eighth-order free energy expansion in polarization powers. Due to multiple potential wells, the height and position of which are temperature- and strain-dependent, the energy profiles of CuInP$_2$S$_6$ can flatten in the vicinity of the non-zero polarization states. Thereby we reveal an unusually strong effect of a mismatch strain on the out-of-plane polarization reversal, hysteresis loops shape, dielectric susceptibility, and piezoelectric response of CIPS films. In particular, by varying the sign of the mismatch strain and its magnitude in a fairly narrow range, quasi-static hysteresis-less paraelectric curves can transform into double, triple, and other types of pinched and single hysteresis loops. The strain effect on the polarization reversal is opposite, i.e., "anomalous", in comparison with many other ferroelectric films, for which the out-of-plane remanent polarization and coercive field increase strongly for tensile strains, and decrease or vanish for compressive strains. For definite values of temperature and mismatch strain, the low-frequency hysteresis loops of polarization may exhibit negative slope in a relatively narrow range of external field amplitude and frequency. The low-frequency susceptibility hysteresis loops, which correspond to the negative slope of polarization loops, contain only positive values, which can be giant in the entire range of field changes. Corresponding piezo-response also reaches giant values, being maximal near coercive fields.

---

[*] Corresponding author, e-mail: anna.n.morozovska@gmail.com

[†] Corresponding author, e-mail: vysochanskii@gmail.com

[‡] corresponding author, e-mail: sergei2@utk.edu

## I. INTRODUCTION

The strain-tunable reversible spontaneous polarization of low-dimensional ferroics materials, including ferroelectrics, ferrielectrics, and antiferroelectrics [1] is of great fundamental interest and are directly defined by the interaction of lattice soft phonons with local atomic potential. Since the strain-tunable reversible polar properties are very attractive for advanced applications in nanoelectronics, sensorics, and 3D multibit memory technologies, strain- and temperature-induced tunability in classical ferroelectric materials have been extensively explored.

Recently, ferroelectricity has been revealed in layered 2D Van der Waals (**V-d-W**) materials, such as $CuInP_2(S,Se)_6$ monolayers, thin films and nanoflakes [2, 3]. The most important aspect of the physics of uniaxial ferrielectrics $CuInP_2(S,Se)_6$ [4, 5], compared to many other uniaxial ferroelectrics, is the existence of more than two potential wells [6], which leads to unique features of their polarization reversal, and, in particular, are responsible for strain-tunable multiple polar states. Due to multiple potential wells, which height and position are and temperature- and strain-dependent, the energy profiles of $CuInP_2(S,Se)_6$ can flatten in the vicinity of the non-zero polarization states. This differs $CuInP_2(S,Se)_6$ from classical ferroelectric materials with the first or second order ferroelectric-paraelectric phase transition, which potential energy profiles can be shallow or flat near the transition point only, corresponding to zero spontaneous polarization. The flatten energy states, in turn, can give rise to enhanced properties associated with polarization dynamics in the vicinity of the nonzero polarization states with the flat potential relief.

The spontaneous polarization of crystalline $CuInP_2S_6$ (**CIPS**) is directed normally to its structural layers as a result of antiparallel shifts of the $Cu^+$ and $In^{3+}$ cations from the middle of the layers [7, 8]. The $Cu^+$ cations flip in their two-well local potential with the between $Cu^{up}$ and $Cu^{down}$ positions with a temperature increase and populate the positions with equal probability above the temperature of the transition from the polar ferrielectric to the nonpolar paraelectric phase. The $In^{3+}$ cations also contribute to the polar ordering of the CIPS crystal lattice, and the shift of the $In^{3+}$ cations in their local potential is opposite to the deviation of $Cu^+$ cations from the middle of CIPS structural layers. **Fig. 1(a)** shows the formation of so-called *low polarization* states in CIPS with the spontaneous polarization $P_S$~(4 – 5) $\mu C/cm^2$, which are studied in this work. Note that **Fig. 1(a)** illustrates only polar and AFE modes involving the shift of $Cu^+$ cations. In a real CIPS structure the mode of $Cu^+$ cations interplays with oppositely polarized polar mode of $In^{3+}$ cations, which leads to the *ferrielectric* ground state. The ferroelectric state with a *high polarization* ~(9 – 10) μC/cm² can be induced only by a very high electric field that returns the $In^{3+}$ cations into the middle of the structural layer, or even into the direction of $Cu^+$ cations shift.

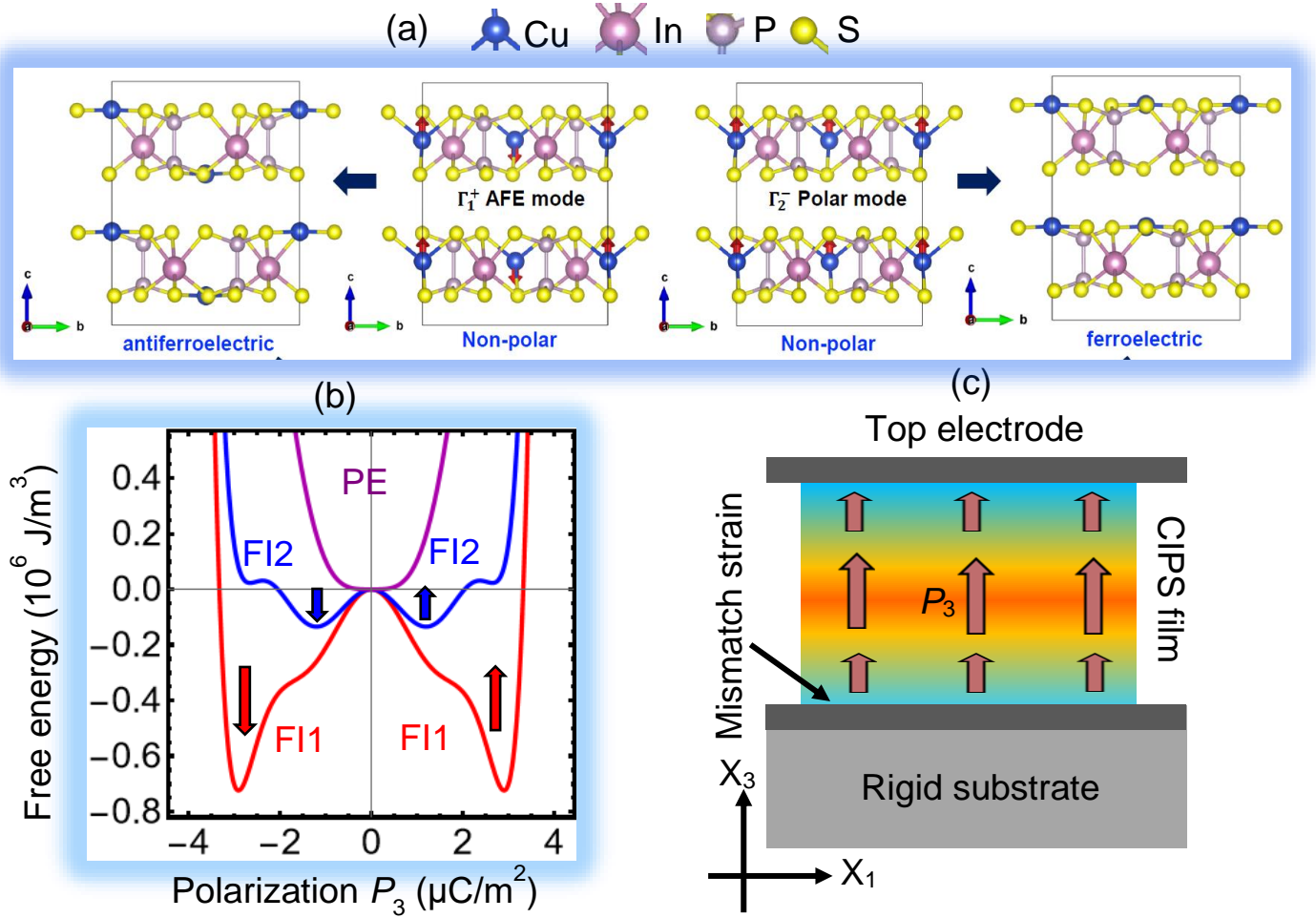

**FIGURE 1**. **(a)** Structural models of CIPS with different polar phases. The non-polar configuration is transformed to a ferroelectric one via the polar (ferroelectric) mode $\Gamma_2^-$. The $\Gamma_1^+$ mode, which is antiferroelectric in nature, leads to the antiferroelectric (AFE) phase. **(b)** Up" and "down" ferrielectric states with big (FI1) and small (FI2) amplitudes of the out-of-plane polarization, shown by the red and blue arrows. **(c)** A thin epitaxial CIPS film sandwiched between ideally-conductive electrodes and clamped on a rigid substrate. Arrows show the direction of the spontaneous polarization.

In pseudospin formalism, the polar ordering of CIPS can be described by the Ising model with spins $s = ½$ and $S = 1$, and a mixed anisotropy of the local crystal field [9], wherein, the spins $\vec{s}$ with projections + ½ and -½ can be associated with the local dipoles induced by $Cu^+$ cations and $P_2S_6$ anion complexes, and spins $\vec{S}$ with projections +1, 0, -1 can be related with the local dipoles induced by $In^{3+}$ cations and $P_2S_6$ anion complexes. In the Landau-Ginzburg-Devonshire (**LGD**) mean-field approximation [10, 11], the presence of two types of cationic sublattices in ferrielectrics is described by polar and antipolar order parameters, *P* and *A*, respectively. They formally correspond to projections $\vec{S} = \pm 1$ in the polar state, and $\vec{S} = 0$ in the non-polar state.

Despite the antipolar order parameter *A* cannot be directly measured, a complete LGD thermodynamic potential for a ferrielectric with the first order phase transition contains even (2-nd, 4-th and 6-th) powers of *P* and *A*, and the biquadratic coupling term between them, $A^2 P^2$. As it was shown in Ref.[11], the biquadratic coupling term induces the term proportional to $P^8$ in the LGD thermodynamic potential for *P*. Using the LGD thermodynamic potential with eighth powers of *P*, we predicted a temperature – stress phase diagram [11, 12, 13], containing the paraelectric phase and two low-polarization ferrielectric states with smaller (~1 µC/cm$^2$) and bigger (~4 µC/cm$^2$) amplitudes of the spontaneous polarization [see **Fig. 1(b)**].

During the polarization reversal in CIPS, due to the antiparallel and antiphase displacement of the ferroactive Cu$^+$ and In$^{3+}$ cations relative to the P$_2$S$_6$ sublattice [8], an ionic charge transfer occurs synchronously [14] or asynchronously [15] with the displacement of Cu$^+$ ions. In this case, the Cu$^+$ cations move between the two allowed positions inside the S$_6$-octahedra in the same P$_2$S$_6$ layer [16], until Cu$^+$ leaves its P$_2$S$_6$ layer and interacts with S$^-$ in the frame of the S$_6$-octahedrons of the neighboring P$_2$S$_6$ layer through the V-d-W gap [14]. From the theoretical standpoint, the polarization dynamics in V-d-W and related ferroelectric materials is governed by an effective multi-well free energy landscape of the order parameter [10 - 13, 17]. The LGD approach [18, 19, 20] can be applied to describe the polarization reversal inside the P$_2$S$_6$ layer in CIPS [21] and related materials. LGD is less efficient in describing the contribution of interlayer polarization to the CIPS V-d-W gap due to their direct relationship with ionic transfer (see, e.g., Ref.[22]). However, a proper elaboration of the LGD approach made it suitable for the description of ferro-ionic [23, 24] and antiferro-ionic [25, 26] interactions.

Here we study strain-induced transitions of polarization reversal in thin strained CIPS films covered by ideally-conductive electrodes [see **Fig. 1(c)**] using the LGD approach. We reveal an unusually strong influence of the mismatch strain on the out-of-plane polarization reversal, hysteresis loops shape, dielectric susceptibility, and piezoelectric response of the films. The original part of this work contains the physical description of the problem (**Section II**), analysis the strain-induced transitions of a polarization reversal, and hysteresis loop changes in the CIPS films (**Section III**). **Section IV** summarizes the obtained results and potential outcomes. **Supplementary Materials** elaborate on a mathematical formulation of the problem, a table of material parameters, a description of methods, and numerical algorithms.

## II. PROBLEM FORMULATION

Here we consider an epitaxial thin CIPS film sandwiched between ideally-conductive electrodes, and clamped on a thick rigid substrate [see **Fig. 1(c)**]. Pink arrows show the out-of-plane ferroelectric polarization $P_3$, directed along X$_3$-axis. The perfect electric contact between the film and ideal electrodes provides an effective screening of the out-of-plane polarization by the electrodes and prevents the domain formation.

Within the continuous media approximation and LGD approach, the value and orientation of the spontaneous polarization $P_i$ in thin ferroelectric films is controlled by the temperature $T$ and mismatch strain $u_m$. The strain $u_m$ originates from the film-substrate lattice constants mismatch and exists entire the film depth [27, 28, 29], because the film thickness is regarded smaller than the critical thickness of misfit dislocations appearance. For the validity of the continuum media approximation, the film thickness is regarded to be much bigger than the lattice constant. As a rule, the conditions are valid in the film thickness range (5 – 50) nm.

It has been shown in Refs.[11 - 13] that the LGD free energy density of CIPS, $g_{LGD}$, has four potential wells at $\vec{E} = 0$. The density $g_{LGD}$ includes the Landau-Devonshire expansion in even powers of the polarization $P_3$ (up to the eighth power), the Ginzburg gradient energy, the elastic and electrostriction energies. These are listed in **Appendix A** of **Supplementary Materials** [30]. The dynamics of polarization $P_3$, piezocoefficient $d_{33}$ and dielectric susceptibility $\chi_{33}$ in an external field $E_3$ follows from the time-dependent LGD equations, which have the form:

$$\Gamma\frac{\partial P_3}{\partial t} + \left[\alpha - 2\sigma_i(Q_{i3} + W_{ij3}\sigma_j)\right]P_3 + (\beta - 4Z_{i33}\sigma_i)P_3^3 + \gamma P_3^5 + \delta P_3^7 - g_{33kl}\frac{\partial^2 P_3}{\partial x_k \partial x_l} = E_3, \quad (1a)$$

$$\Gamma\frac{\partial \chi_{33}}{\partial t} + \left[\alpha - 2\sigma_i(Q_{i3} + W_{ij3}\sigma_j) + 3(\beta - 4Z_{i33}\sigma_i)P_3^2 + 5\gamma P_3^4 + 7\delta P_3^6\right]\chi_{33} - g_{33kl}\frac{\partial^2 \chi_{33}}{\partial x_k \partial x_l} = 1, \quad (1b)$$

$$\Gamma\frac{\partial d_{33}}{\partial t} + \left[\alpha - 2\sigma_i(Q_{i3} + W_{ij3}\sigma_j) + 3(\beta - 4Z_{i33}\sigma_i)P_3^2 + 5\gamma P_3^4 + 7\delta P_3^6\right]d_{33} - g_{33kl}\frac{\partial^2 d_{33}}{\partial x_k \partial x_l} =$$
$$(2Q_{33} + 2W_{i33}\sigma_i)P_3 + 4Z_{333}P_3^3. \quad (1c)$$

Here, $\Gamma$ is the Khalatnikov kinetic coefficient [31]. The coefficient $\alpha$ depends linearly on the temperature $T$, $\alpha(T) = \alpha_T(T - T_C)$, where $T_C$ is the Curie temperature of a bulk ferrielectric. The coefficients $\beta$, $\gamma$, and $\delta$ in Eq.(1) are temperature independent. The values $\sigma_i$ denote diagonal components of a stress tensor in the Voigt notation, and the subscripts $i, j = 1 - 6$. The values $Q_{i3}$, $Z_{i33}$, and $W_{ij3}$ denote the components of a single linear and two nonlinear electrostriction strain tensors in the Voigt notation, respectively [32, 33]. The values $g_{33kl}$ are polarization gradient coefficients in the matrix notation and the subscripts $k, l = 1 - 3$. Note that the contribution of nonlinear electrostriction $W_{ij3}$ was not considered in Ref.[12]. Corresponding analytical expressions for the coefficients $\alpha^*$, $\beta^*$, $\gamma^*$, and $\delta^*$, renormalized by the mismatch strain, are derived in this work and listed in **Appendix A**.

The boundary condition for $P_3$ at the film surfaces S is "natural", i.e., $g_{33kl}n_k\frac{\partial P_3}{\partial x_l}\Big|_S = 0$, where $\vec{n}$ is the outer normal to the surface. The value $E_3$ in Eq.(1a) is an electric field component co-directed with the polarization $P_3$. In a general case, it is a superposition of external and depolarization fields. In the considered case of ideal screening, the depolarization field, as well as domain formation, are absent. This is because of ideal screening, which makes the solutions corresponding to constant $P_3$ most energetically-favorable. Hence, polarization gradient is absent in Eqs.(1a)-(1c). To analyze a quasi-static polarization reversal, we assume that the period, $2\pi/\omega$, of the sinusoidal external field is very small in comparison with the Landau-Khalatnikov relaxation time, $\tau = \Gamma/|\alpha|$.

Modified Hooke's law relating elastic strains $u_i$ and stresses $\sigma_j$, is obtained from the relation $u_i = -\partial g_{LGD}/\partial \sigma_i$:

$$u_i = s_{ij}\sigma_j + Q_{i3}P_3^2 + Z_{i33}P_3^4 + W_{ij3}\sigma_j P_3^2. \qquad (2)$$

For the considered CIPS film the following relations are valid for homogeneous stress and strain components: $\sigma_3 = \sigma_4 = \sigma_5 = \sigma_6 = 0$, $u_1 = u_2 = u_m$ and $u_4 = u_5 = u_6 = 0$.

The values of $T_C$, $\alpha_T$, β, $\gamma$, δ, $Q_{i3}$, $W_{ij3}$, and $Z_{i33}$ have been derived in Refs.[12, 13] from the fitting of temperature-dependent experimental data for the dielectric permittivity [34, 35], spontaneous polarization [36], and lattice constants [37] as a function of hydrostatic pressure. Elastic compliances $s_{ij}$ were calculated from ultrasound velocity measurements [38, 39]. The CIPS parameters used in our calculations are listed in **Table SI** in **Appendix A**.

## III RESULTS AND DISCUSSION

### A. Quasi-static behavior of polarization reversal

The strain effect on the polarization reversal in "normal" ferroelectric films with $Q_{33} > 0$, $Q_{23} < 0$, and $Q_{13} < 0$ typically leads to a strong increase of the out-of-plane spontaneous polarization and coercive field for tensile strains, and their strong decrease or disappearance for compressive strains (see e.g., Ref. [27, 29]). Due to the "inverted" signs of the linear electrostriction coupling coefficients ($Q_{33} < 0$, $Q_{23} > 0$, and $Q_{13} > 0$) as well as due to the strongly negative and temperature-dependent nonlinear electrostriction coupling coefficients ($Z_{i33} < 0$ and $W_{ij3} < 0$) for CIPS (see **Table SI**), we expect that the strain effect on the CIPS film remanent polarization and coercive field can differ in comparison with films of classical ferroelectrics. Below we show and analyze two types of the phase diagrams: the "static" diagram (**Fig. 2**), which defines the thermodynamically stable polarization ground state, and the "dynamic" diagrams (**Fig. 3**) representing the quasi-static polarization dynamics.

The dependence of the spontaneous polarization $P_s$ on temperature $T$ and mismatch strain $u_m$ is shown in **Fig. 2(a)**. The diagram, which is consistent with a central part of the diagram 4(b) in Ref.[12], was calculated by a conventional numerical minimization of the free energy (S.1) listed in **Appendix A**. The color scale in the diagram shows the absolute value of $P_s$ in the deepest potential well of the LGD free energy. Here, a wedge-like region of the paraelectric (**PE**) phase, which is stable at $T > 300$ K, meets with the two ferrielectric states, **FI1** and **FI2**, which correspond to big and small amplitudes of the out-of-plane spontaneous polarizations $P_3^\pm$, respectively. The schematics of a potential relief in the PE phase, FI1 and FI2 states are shown in **Fig. 2(b)**.

The PE-FI2 boundary is the line of second order phase transition at tensile strains $u_m > 0$, and PE-FI1 boundary is the first order transition at compressive strains $u_m < 0$. The first order transition stipulates the coexistence of the FI1 state and the PE phase for tensile strains, marked as the FI1+PE region. The critical end point (**CEP**), where the second order PE – FI2 phase transition line terminates, is shown by a black circle; and the bicritical end point (**BEP**), where the first order isostructural phase transition line between two ferrielectric states (**FI1** and **FI2**) with different amplitudes of spontaneous polarization terminates, is shown by a white circle.

The anomalous feature of the diagram is that the high-polarization FI1 state exists at compressive strains $u_m < 0$. The state does not vanish for a tensile strain $u_m > 0$; instead, it continuously transforms in the small-polarization FI2 state at $u_m > 0$, and eventually undergoes the second order phase transition to the PE phase at the dotted line. The situation, shown in **Fig. 2(a)** for $u_m > 0$, is anomalous for the most uniaxial and multiaxial ferroelectric films, where the out-of-plane polarization is absent or very small at $u_m > 0$, and the region of FE c-phase vanishes or significantly constricts for $u_m > 0$ [27]. This analysis also illustrates that the strain effects in CIPS and similar materials can result in flattening of the free energy profile around nonzero polarization values, unlike classical ferroelectrics. Therefore, these regions can be prone to emergence of the giant electromechanical and dielectric responses, as will be explored below.

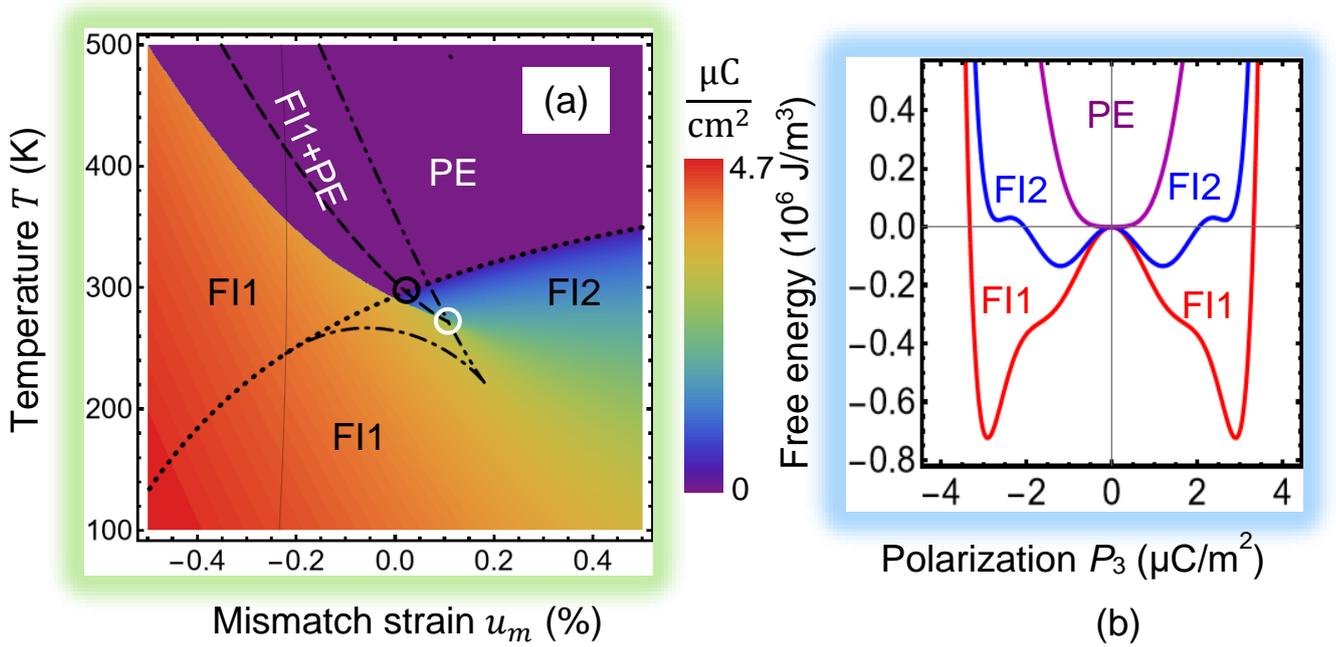

**FIGURE 2**. **(a)** The dependence of the spontaneous polarization $P_S$ on temperature $T$ and mismatch strain $u_m$. PE is the paraelectric phase, FI1 and FI2 are the ferrielectric states. CEP and BEP are the critical and bicritical end points, marked by the black and white circles, respectively. Color scale is the absolute value of $P_S$ in the deepest potential well of the LGD free energy. **(b)**. Schematics of a potential relief in the PE phase (the dark-violet curve), FI1 (the red curve) and FI2 (the blue curve) states.

The diagram in **Fig. 3(a)** illustrates the influence of temperature $T$ and mismatch strain $u_m$ on the shape of static curves and quasi-static hysteresis loops, $P_3(E)$. **Fig. 3(b)** is a zoomed central region of **Fig. 1(c).** The diagrams were calculated from dynamic Eqs.(1) at $\omega = 0$ (for the static curves) and $\omega\tau \ll 10^{-4}$ (for quasi-static hysteresis loops). The diagrams contain a red region of paraelectric curves (**PC**); an orange region of double loops (**DL**); yellow and olive regions of triple loops of the first (**TL-I**) and second (**TL-II**) types; light-green, brown and dark-green regions of pinched loops of the first (**PL-I**), second (**PL-II**) and third (**PL-III**) types; cyan, dark-cyan and blue regions of single loops of the first (**SL-I**), second (**SL-II**) and third (**SL-III**) types. The classification takes into account the shape of the quasi-static loops and the structure of the static curves, which are shown in **Fig. 3(c)** by red solid curves and thin black dashes, respectively. Note, that the ten types of curves distinguished in thin strained CIPS films and shown in **Fig. 3(c)**, are bigger than eight types of curves distinguished in CIPS nanoparticles in Ref. [13]. Equations describing different curves, which separate different regions in **Fig. 3** are listed in **Appendix B**.

The yellow and olive regions of TL-I and TL-II shown in **Fig. 3(b)** are located inside the blue region of FI2 phase shown in **Fig. 2(a).** TLs fill the region inside the dotted, dashed, and dot-dot-dashed curves. The CEP and BEP are 2 vertices of the triangle. The region of PL-I is located above the dashed curve in the PE and FI1 coexisting region. Thin PL-II and PL-III regions are stable near the diffuse boundary between the FI2 and FI1 states. As follows from the structure of the static curves in **Fig. 3(c)**, TLs and PLs can be imagined as the result of superposition of a "central" single loop (corresponding to the switching of the "small" polarization in the FI2 state) and a double loop (corresponding to the switching of a "larger" polarization in the FI2 state). The coexistence of these two polarizations in the four-well FI2 state determines the stability of the TLs shown in **Figs. 3**.

The cyan region of SL-I and dark-cyan region of SL-II are separated by the dotted curves in **Figs. 3(a)** and **3(b)**. The SL-I and SL-II regions are located inside the FI1 state in **Fig. 2(a)**. The SL-I region is also stable in the PE and FI1 coexistence region. Bid blue regions of SL-III, shown **Fig. 3(a)** and **3(b),** are reentrant, as located inside the regions of FI1 and FI2 absolute stability in **Figs. 2(a)**. The absolute stability of the SL-III appears when the four-well potential transforms into the two-well potential.

The regions of PC and DL, separated by a dot-dot-dashed curve in **Fig. 3(b)**, are located inside the PE phase region shown in **Fig. 2(a)**. These loops are related to the first order phase transition induced by the electric field.

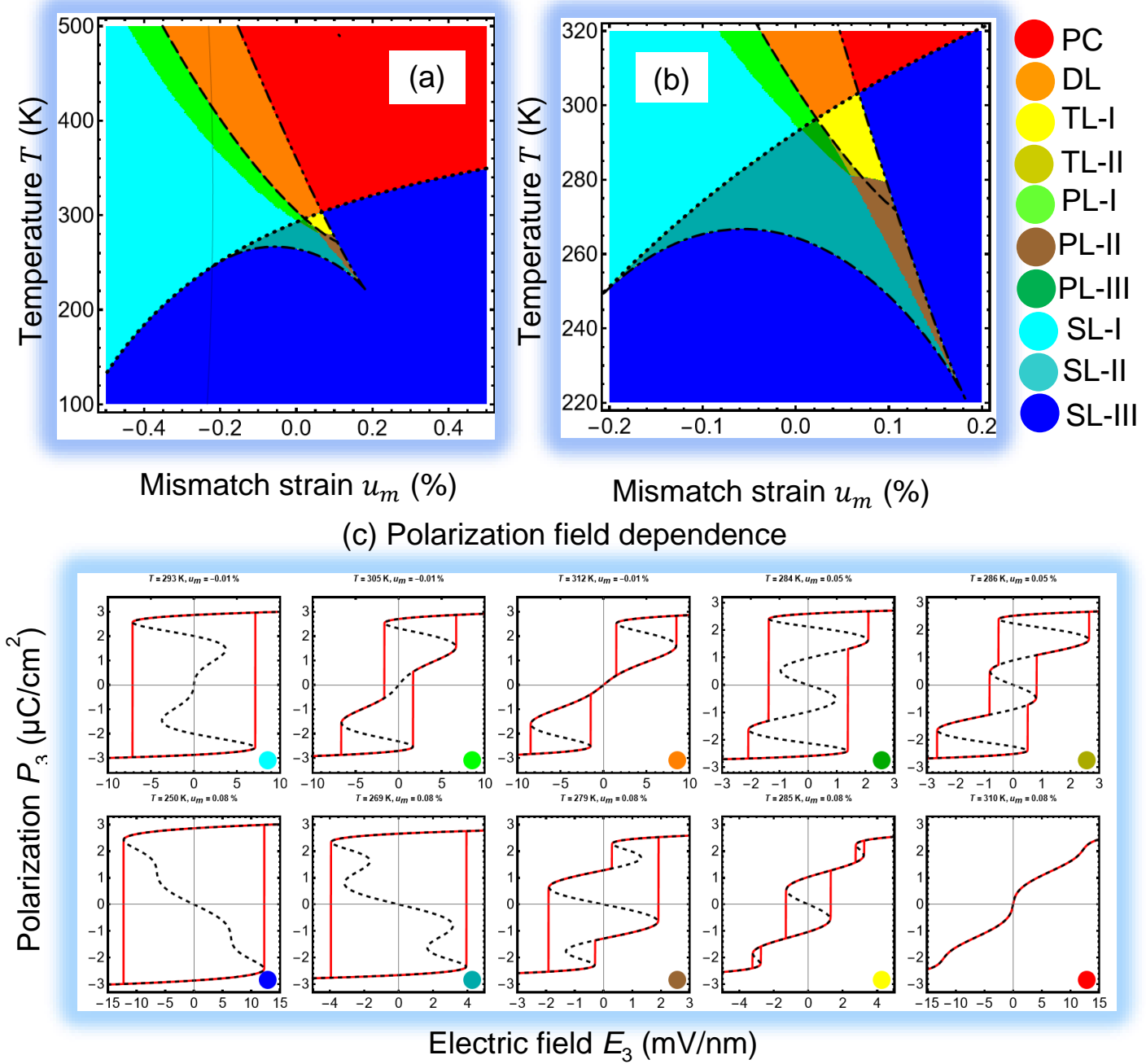

**FIGURE 3**. **(a)** The shape of quasi-static hysteresis loops, $P_3(E)$, calculated as a function of temperature $T$ and mismatch strain $u_m$. **(b)** A zoomed central region of the diagram **(a)**. Color coding in the diagrams (a-b): red is paraelectric curves (PC), orange is double loops (DL); yellow and olive are triple loops of the first (TL-I) and second (TL-II) types; light-green, brown and dark-green are pinched loops of the first (PL-I), second (PL-II) and third (PL-III) types; cyan, dark-cyan and blue are single loops of the first (SL-I), second (SL-II) and third (SL-III) types. **(c)** Electric field dependence of the polarization $P_3$, calculated for CIPS thin films, different temperature $T$ and mismatch strain $u_m$ listed above each graph. Black dashed curves are static dependences ($\omega = 0$); red solid loops are quasi-static hysteresis loops calculated for dimensionless frequency $\omega\tau = 10^{-5}$. Colors of the circles correspond to the diagram **(b)**. CIPS parameters are listed in **Table I**.

As anticipated, far from all features of the static curves are present on the quasi-static loops (compare red solid loops and black dashed curves in **Fig. 3(c)**). In particular, unstable parts of the static curves, which correspond to the negative susceptibility in **Fig. 4(a)**, as well as those stable parts of the static curves, which are located inside the SL, are never "observed" at the quasi-static loops. At the same time, static curves, and quasi-static loops of dielectric susceptibility $\chi_{33}$ and piezoelectric coefficient $d_{33}$, shown in **Fig. 4(a)** and **4(b)**, reflect all features of the polarization static curves and quasi-static loops, respectively. In particular, the static curves of $\chi_{33}$ and $d_{33}$ have divergencies at coercive fields and/or sharp jumps, their position coinciding with the jumps on polarization static curves. The quasi-static loops of $\chi_{33}$ and $d_{33}$ almost coincide with the static curves, except for the unstable parts and divergencies at coercive fields. The divergencies transform into sharp maxima, their height decreasing with increase in frequency $\omega$. The quasi-static loops of $\chi_{33}$ and $d_{33}$ contain two, four or six features (jumps and/or sharp maxima), which correspond to different types of SLs, PLs or TLs, respectively. Sometimes some of these features are hardly seen on the polarization loops, especially under the transition from one type of the loop to another. However, the features are much better seen on the loops of susceptibility and piezocoefficient, and thus the simultaneous measurements of the polarization, susceptibility and piezoelectric response can help to verify our theoretical predictions.

An important peculiarity of **Figs. 3** and **4** is the strong influence of the elastic strain on the features of quasi-static polarization reversal in thin CIPS films, namely the variability of hysteresis loops shapes, and particularly the existence of the PL and TL regions at $u_m$ <0.2%. This is due to the specific structure of the static curves, which is determined by the 8-th order LGD potential. The absence of PC and DL at compressive strains $u_m$ <0 and $T < 300$ K originates from the anomalous temperature-dependence and "inverted" sign of the CIPS linear and nonlinear electrostriction coupling coefficients.

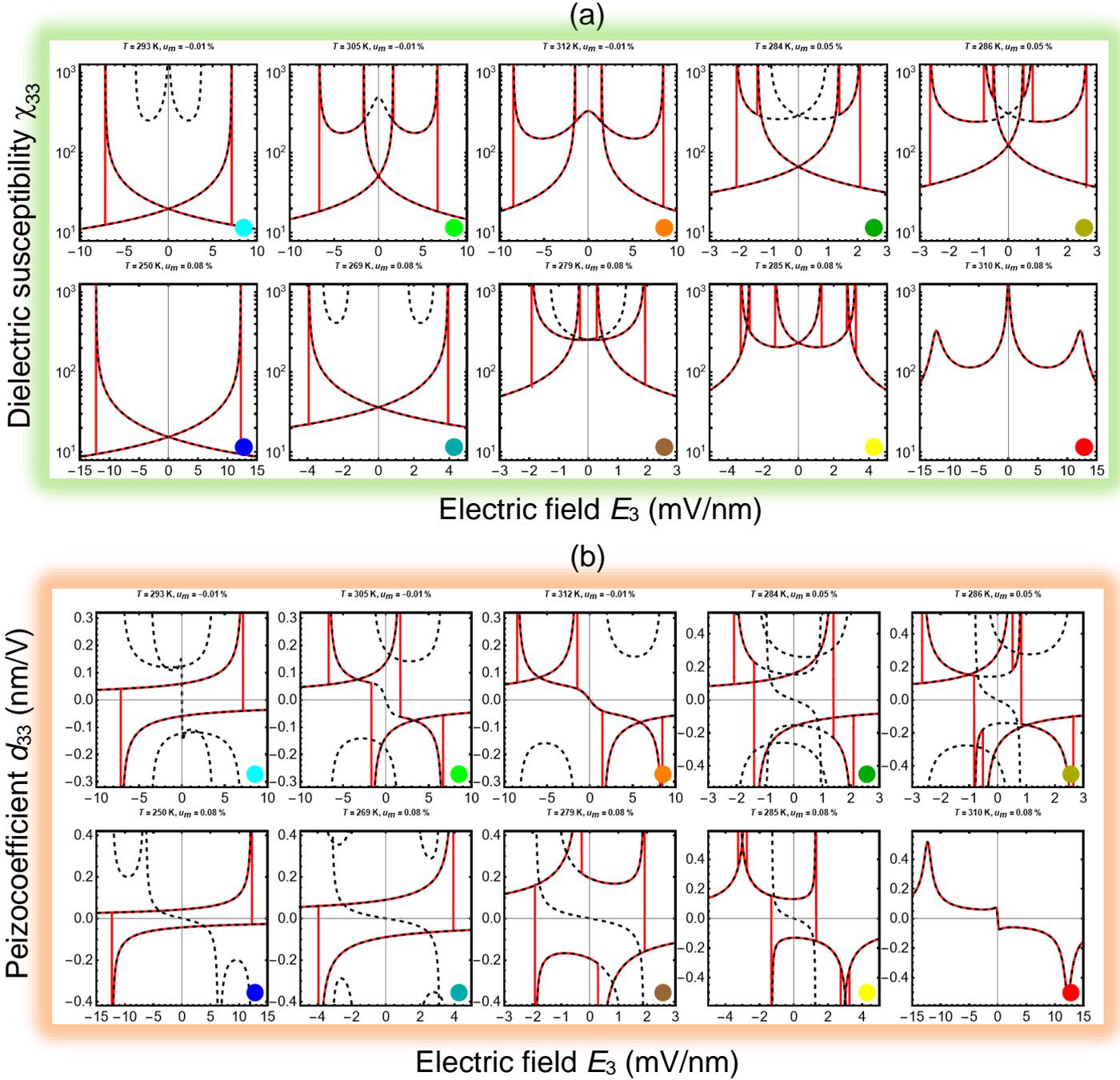

**FIGURE 4**. Electric field dependence of the dielectric susceptibility $\chi_{33}$ **(a)** and piezocoefficient $d_{33}$ **(b)** calculated for CIPS thin films, different temperature $T$ and mismatch strain $u_m$ listed above each graph. Black dashed curves are static dependences ($\omega = 0$); red solid loops are quasi-static hysteresis loops calculated for dimensionless frequency $\omega\tau = 10^{-5}$. Colors of the circles correspond to the diagram in **Fig. 3(b)**.

### B. Low-frequency behavior of polarization reversal

The low-frequency hysteresis loops of polarization $P_3$, piezoelectric coefficient $d_{33}$ and relative dielectric susceptibility $\chi_{33}$, calculated for frequencies $\omega\tau \leq 10^{-3}$, look very similar to the quasi-static loops shown in **Figs. 3(c)** and **4**. However, polarization hysteresis loops calculated in the frequency range $10^{-2} \leq \omega\tau \leq 5 \cdot 10^{-1}$, relatively narrow range of external field amplitude $E_0$, and definite ranges of temperature and mismatch strain, may have a negative slope [see e.g., red, blue and green loops in **Figs. 5(a)** and **5(d)**]. In particular, the negative slope can be observed for those hysteresis loops, which static curves have three unstable regions and two touch points with P-axis [see e.g., black dashed curves in **Figs. 5(a)** and **5(d)**].

One should expect that the negative slope of hysteresis loops may correspond to the negative relative dielectric susceptibility, which is transient in the considered case ($\omega\tau > 10^{-2}$). Indeed, the metastable parts of dashed black curves in **Figs. 5(a)** and **5(d)** correspond to negative values of static susceptibility. A point-by-point derivative, $dP_3/dE_3$, calculated numerically using the points of solid curves, is negative in the region of negative slope, allowing to speculate about the possibility of transient negative capacitance of the film. However, the "true" dielectric susceptibility, $\chi_{33}$, calculated from Eq.(1b) for parameters, which correspond to the solid loops in the plots **5(a)** and **5(d),** is positive as anticipated from thermodynamic equilibrium principle. This positive susceptibility is shown in **Figs. 5(b)** and **5(e)**.

Interestingly, the susceptibility and piezoelectric coefficient hysteresis loops, shown in **Figs. 5(b, c)** and **5(e, f)**, which correspond to the green polarization loop with negative slope in **Fig. 5(a)** and red polarization loop with negative slope in **Fig. 5(d)**, reach giant values (such as $\chi_{33}$ >10$^4$ and $d_{33}$ >5nm/V) entire the range of field changes, and are maximal for small fields. The susceptibility and piezocoefficient loops, calculated for higher $E_0$ and shown by the blue and red loops in **Fig. 5(b)** and **5(c)**, as well as by blue and green loops in **Fig. 5(d)** and **5(f)**, are significantly smaller and reach maximal values in a relatively wide vicinity of the coercive field. In contrast to the single hysteresis loops in many uniaxial ferroelectrics described by a two-well or three-well LGD potential (corresponding to the 2-4 powers or 2-4-6 powers of polarization series, respectively), the loops of dielectric susceptibility and piezocoefficient for CIPS films have additional maxima near coercive fields originated from its four-well LGD potential (corresponding to the 2-4-6-8 powers of polarization series).

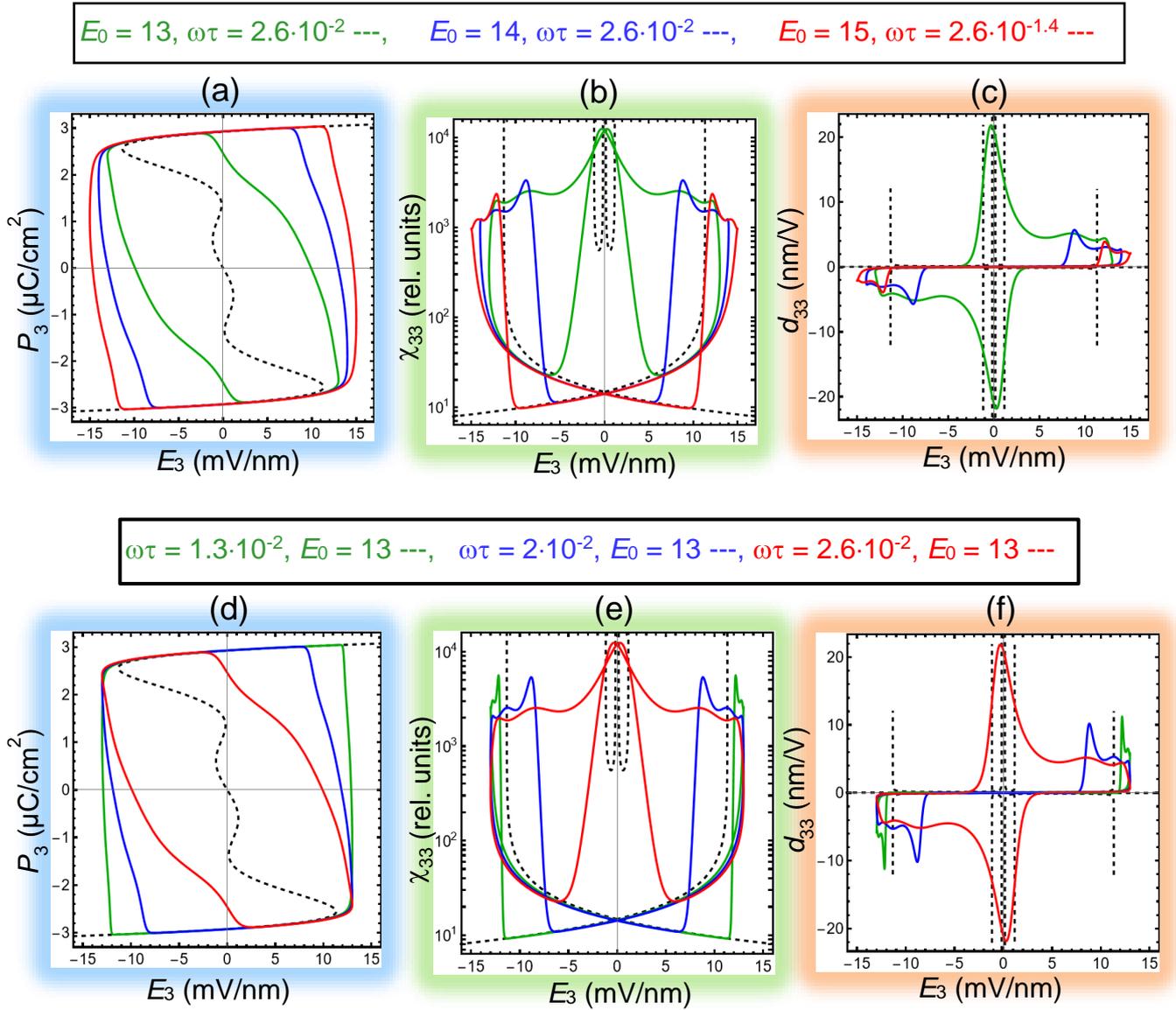

**FIGURE 5.** Electric field dependence of the polarization $P_3$ **(a, d)**, dielectric susceptibility $\chi_{33}$ **(b, e)**, and piezocoefficient $d_{33}$ **(c, f)** calculated for CIPS thin films. Black dashed curves are static dependences ($\omega = 0$). Green, blue, and red solid curves are dynamic hysteresis loops calculated for different amplitudes $E_0 = (13 - 15)$ mV/nm of applied field and the same frequency $\omega\tau = 2.6 \cdot 10^{-2}$ **(a, b, c)**; or for different frequencies $\omega\tau = (1.3 - 2.6) \cdot 10^{-2}$ of applied field and the same amplitude $E_0 = 13$ mV/nm **(d, e, f)**. CIPS parameters are listed in **Table I**, $T = 276$ K and $u_m = 0.01\%$.

The giant values $\chi_{33}$ and $d_{33}$, which correspond to the polarization hysteresis loops with negative slope, presents further interest and can be useful for practical applications. Hence, we proceed to determine the ranges of temperature $T$ and mismatch strain $u_m$ for which the negative slope can be observed. Using analytical conditions listed in **Appendix B** and numerical algorithm described in **Appendix C**, we determined the intervals of dimensionless frequency $\omega\tau$ and external field amplitude $E_0$ for which the polarization hysteresis loops have negative slope for definite values of $T$ and $u_m$. Corresponding dependence of $T$ on $u_m$ is shown by a thick dark-green curve surrounded by a diffuse green region in **Fig. 6(a)**. The green curve is calculated from Eqs.(S.10a) and (S.10b), which are listed in **Appendix B**. The Eqs.(S.10) determine the conditions of zero coercive field for internal unstable parts of the static curves shown in, e.g., **Figs. 5(a)** and **5(d)**. The diagrams, shown in **Figs. 6(c)**, **6(d)** and **6(e)**, show the ranges of $\omega\tau$ and $E_0$ for which either the partial switching of polarization, or hysteresis loops with a negative slope, or hysteresis loops with a positive slope are stable for $T \cong (270 - 275)$ K, negative, almost zero or positive $u_m$, respectively. In the diagrams, the dark-green region of the loops with a negative slope separates the dark-red region of a partial switching from the blue region of the loops with a positive slope. Notably that the $E_0$-width of the dark-green region is very small for $\omega\tau < 10^{-2}$, but strongly increases with $\omega\tau$ increase above $10^{-2}$, indicating on the transient nature of the negative slope effect. For further frequency increase loops with negative slope eventually transform into quasi-circular or elliptic loops.

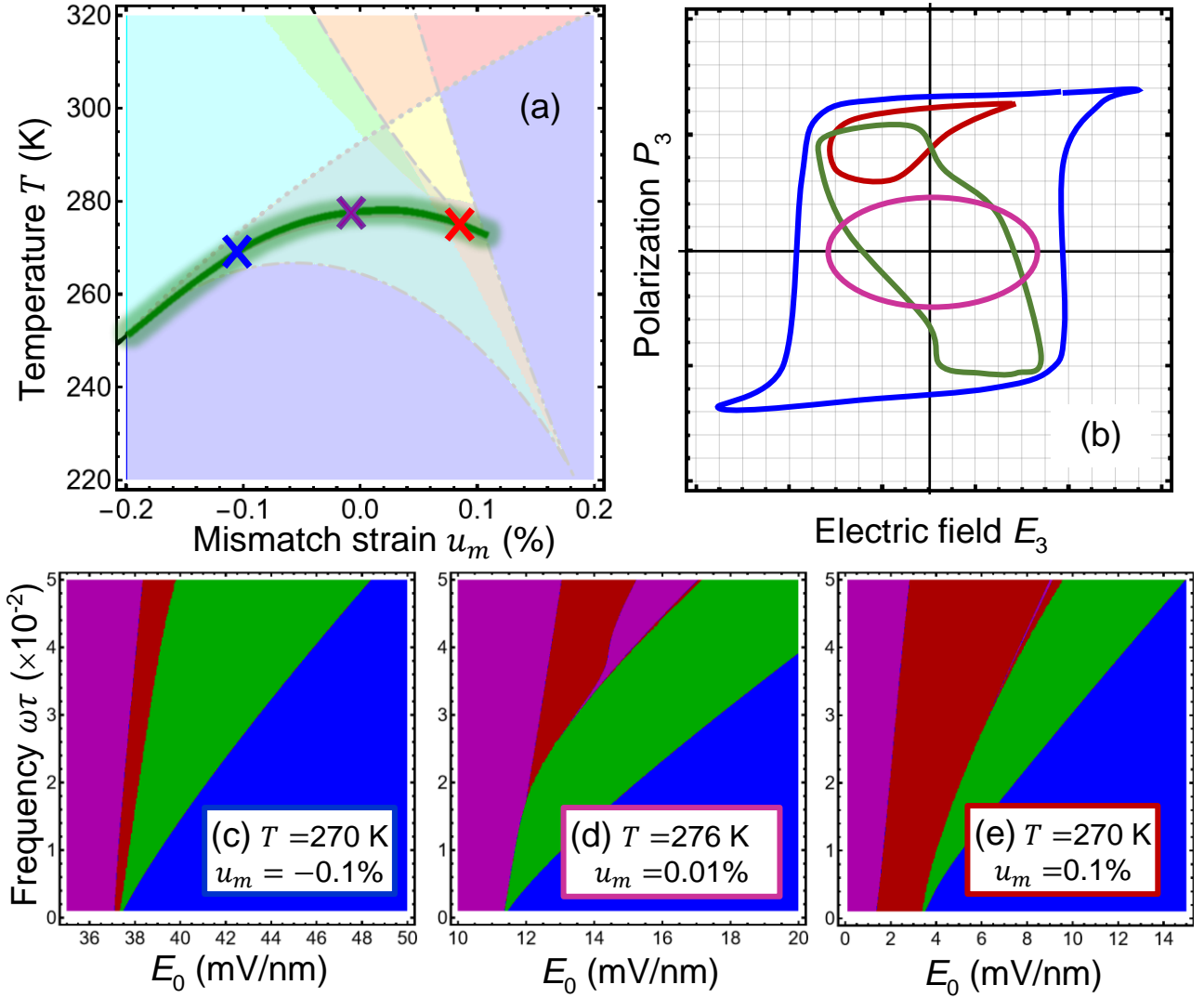

**FIGURE 6.** (a) The ranges of temperature $T$ and mismatch strain $u_m$ for which the negative slope of polarization hysteresis loops can be observed in definite intervals of dimensionless frequency $\omega\tau$ and external field amplitude $E_0$, shown by the diffuse dark-green curve. (b) The schematic image of the polarization partial switching (dark-red color), the elliptic loop of polarization (dark-magenta color), the polarization hysteresis loop with a negative slope

(dark-green color), and the polarization hysteresis loop with a positive slope (blue color). **(c, d, e)** The ranges of $\omega\tau$ and $E_0$ for which the polarization partial switching (dark-red region), elliptic loop hysteresis loops with a negative slope (dark-green region), elliptic and quasi-circular loops (dark-magenta region), and hysteresis loops with a positive slope (blue region) are observed. The diagrams **(c) – (e)** are calculated for different $T$ and $u_m$ listed in legends, which also correspond to the blue, violet and red crosses in the diagram **(a)**. CIPS parameters are listed in **Table I**.

### C. Discussion

Now one may ask, how individual are the polarization reversal features found by us for thin strained CIPS films, which are described by the four-well 2-4-6-8-power LGD potential? The most interesting anomalies of the polarization behavior are observed near phase transitions, CEP and BEP points. In **Figure 7**, the top and middle rows show the anomalies of the polarization behavior in a strained CIPS film. The bottom row illustrates the well-known behavior of a uniaxial single-domain bulk ferroelectric described by the sixth order Landau-Devonshire potential, $g_{LD} = \frac{\alpha}{2}P_3^2 + \frac{\beta}{4}P_3^4 + \frac{\gamma}{6}P_3^6$, for the case $\alpha = 0$, $\beta > 0$, and $\gamma > 0$, which corresponds to the "flat" well at the ferroelectric-paraelectric transition point. We emphasize that the sixth order Landau-Devonshire expansion cannot reveal the four-well potential relief responsible for the appearance of the FI1 and FI2 states, as well as triple loops and loops with negative slope. For nonzero $\alpha$, the sixth order expansion can demonstrate a single-, two-, or three-well potential relief with a central well and two side wells.

**Figure 7(a)** shows the case in which the polarization dependence of the CIPS free energy has four equal potential wells (the first on the left plot). Corresponding polarization, dielectric susceptibility and piezocoefficient dependences on electric field are shown in the second on the left and two plots on the right, respectively. Black dashed curves, which contain three unstable regions, are static dependences. Round-shaped solid loops are low-frequency polarization hysteresis loops calculated for smaller (blue) and bigger (red) amplitudes $E_0$ of applied field. The static dependencies of dielectric susceptibility and piezocoefficient have a complex shape with six divergencies. The low-frequency dielectric susceptibility and piezocoefficient loops have anomalous shape with several maxima and self-crossing, which, in fact, are conditioned by the structure of the static dependences.

**Figure 7(b)** shows the case in which the dependence of CIPS film free energy on polarization has two deep potential wells and two inflection points (the first on the left plot). Corresponding polarization hysteresis with a negative slope in shown in the second on the left plot for smaller (blue) and bigger (red) amplitudes $E_0$. The field dependences of dielectric susceptibility and piezocoefficient are shown in the right plots. Black dashed curves, which contain three unstable regions and two touch points with P-axis, are static dependences. The static dependencies of dielectric susceptibility and piezocoefficient look like six very narrow divergent peaks. The low-frequency dielectric susceptibility and piezocoefficient loops have anomalous shape with several maxima and self-crossing, which position are different from the static dependences.

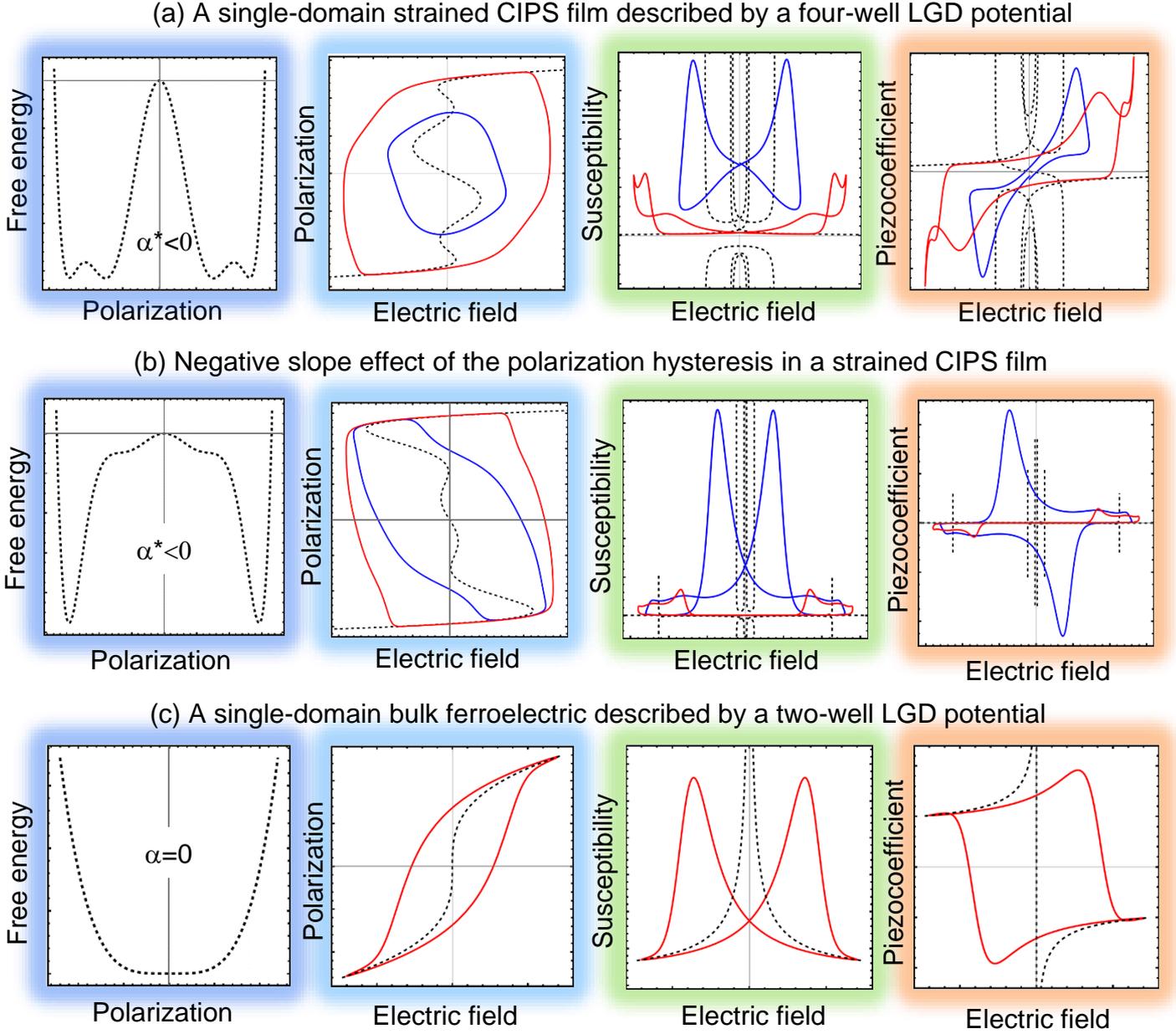

**FIGURE 7**. A single-domain strained CIPS film described by a four-well LGD potential **(a)**, negative slope effect of the polarization hysteresis in the film **(b)**, and a single-domain bulk ferroelectric described by a two-well LGD potential calculated for α=0 **(c)**: free energy dependence on polarization (left), polarization $P_3$, dielectric susceptibility $\chi_{33}$ and piezo-coefficient $d_{33}$ dependence on electric field $E_3$ (right). Black dashed curves are static dependences ($\omega = 0$). Solid curves are dynamic hysteresis loops calculated for smaller (blue) and bigger (red) amplitudes $E_0$ of applied field and the same frequency $\omega\tau \cong 10^{-2}$, where $\tau = \Gamma/|\alpha_T T_C|$.

Both static and low-frequency dependences, shown in **Fig. 7(a)** and **7(b)**, have principal differences from the corresponding dependences of a single-domain bulk ferroelectric described by a flat-well Landau-Devonshire potential in the vicinity of the ferroelectric-paraelectric transition point [see

**Fig. 7(c)**]. Hence, the above features look individual for CIPS and related ferro- or ferrielectric materials, whose free energy profiles can flatten in the vicinity of the nonzero polarization states.

Next question is which of these features will survive in the epitaxial CIPS films covered with non-ideal electrodes or surface screening charges? The answer depends on the so-called effective screening length λ of the non-ideal electrodes, which should be small enough to prevent the domain formation in the film. Our calculations for CIPS, as well as earlier LGD-based calculations for classical first and second order ferroelectric materials [40], show that, as a rule, the single-domain state remains stable for λ≤0.01 nm in a (20 – 50) nm-thick film. The value 0.02 nm corresponds to a semi-metallic electrode, such as $SrRuO_3$ with carrier concentration about $10^{28}$ m$^{-3}$, and the λ-values smaller than 0.01 nm are reached in metallic (e.g., Pt, Ag or Au) electrodes. Thus, in order to verify the effects predicted in this work, one rather needs to provide an ideal electric contact between the epitaxial film and the standard metallic electrodes.

## IV. CONCLUSIONS

Using the eighth-order LGD free energy expansion in polarization powers, we study strain-induced transitions of polarization reversal scenario in CIPS thin films covered by ideally-conductive electrodes. Due to multiple potential wells, which height and position are temperature- and strain-dependent, the energy profiles of CIPS can flatten in the vicinity of nonzero polarization states. This differs CIPS from classical ferroelectric materials with the first or second order ferroelectric-paraelectric phase transitions, which potential energy profiles can be shallow or flat near the transition points only, corresponding to zero spontaneous polarization. Due to the difference, we reveal an unusually strong influence of a mismatch strain on the out-of-plane polarization reversal, hysteresis loops shape, dielectric susceptibility, and piezoelectric response of the strained CIPS films. In particular, by varying the sign of the mismatch strain (from tension to compression), and its magnitude (from zero to 0.2%), quasi-static hysteresis-less paraelectric curves can transform into double, triple, and different types of pinched and single hysteresis loops. The specific shape of the quasi-static hysteresis loops is defined by static dependences of polarization on an applied electric field, referred to as "static curves", which can contain one, two or four unstable parts, related with a multi-well LGD potential of CIPS.

The predicted strain effect on the polarization reversal is opposite, i.e., "anomalous", in comparison with many other ferroelectric films: the out-of-plane remanent polarization enhances strongly and coercive field increases for tensile strains, meanwhile the polarization decreases strongly or even disappears, and hysteresis characteristics worsen significantly for compressive strains. We explain this effect by "inverted" signs of linear and nonlinear electrostriction coupling coefficients of CIPS and their strong temperature dependence.

For definite values of temperature and mismatch strain the low-frequency hysteresis loops of polarization may have a negative slope in the relatively narrow range of external field amplitude and frequency. In particular, the negative slope of hysteresis loops can be observed in the states with a four-well LGD potential, which has two deep wells and either two very shallow wells or two inflection points. The field-width of the negative slope region strongly increases with the frequency increase. For further frequency increase loops with negative slope eventually transform into quasi-circular or elliptic loops, indicating on the transient nature of the negative slope effect, and eventually transforms into quasi-circular or elliptic loops. Notably, the low-frequency susceptibility hysteresis loops, which correspond to the negative slope of polarization loops, contain only positive values, which can be giant (such as $10^4$) in the entire range of field changes, and are maximal at very small fields. Corresponding piezoelectric response also can reach giant values (such as 5 nm/V) in the entire range of field changes, and is maximal near coercive fields.

**Acknowledgements.** S.V.K. was supported by the center for 3D Ferroelectric Microelectronics (3DFeM), an Energy Frontier Research Center funded by the U.S. Department of Energy (DOE), Office of Science, Basic Energy Sciences under Award Number DE-SC0021118.A.N.M. acknowledges support from the National Research Fund of Ukraine (project "Low-dimensional graphene-like transition metal dichalcogenides with controllable polar and electronic properties for advanced nanoelectronics and biomedical applications", grant application 2020.02/0027). E.A.E. acknowledges support from the National Academy of Sciences of Ukraine. Authors are very grateful to Prof. Nicholas Morozovsky (NASU) for useful discussions and valuable suggestions.

**Data availability statement.** Numerical results presented in the work are obtained and visualized using a specialized software, Mathematica 13.1 [41]. The Mathematica notebook, which contain the codes, is available per reasonable request.

**Authors' contribution.** The research idea belongs to A.N.M., Yu.M.V. and S.V.K. A.N.M. formulated the problem, performed analytical calculations, analyzed results, and wrote the manuscript draft. E.A.E. and M.E.Y. wrote codes; and prepared Suppl. Materials. A.G. provided atomic structure models. Yu.M.V. and S.V.K. worked on the results explanation and manuscript improvement. All co-authors discussed the obtained results.

<div align="center">

**SUPPLEMENTARY MATERIALS**

**Appendix A. Landau-Ginzburg-Devonshire approach**

</div>

The density of the CIPS LGD potential, $g_{LGD}$, which includes the Landau-Devonshire expansion in even powers of the polarization $P_3$ up to the eighth power, $g_{LD}$, the Ginzburg gradient energy $g_G$, and the elastic and electrostriction energies, $g_{ES}$, has the form [42, 43]:

$$g_{LGD} = g_{LD} + g_G + g_{ES}, \qquad (S.1a)$$

$$g_{LD} = \frac{\alpha}{2}P_3^2 + \frac{\beta}{4}P_3^4 + \frac{\gamma}{6}P_3^6 + \frac{\delta}{8}P_3^8 - P_3 E_3, \tag{S.1b}$$

$$g_G = g_{33kl}\frac{\partial P_3}{\partial x_k}\frac{\partial P_3}{\partial x_l}, \tag{S.1c}$$

$$g_{ES} = -\frac{s_{ij}}{2}\sigma_i\sigma_j - Q_{i3}\sigma_i P_3^2 - Z_{i33}\sigma_i P_3^4 - W_{ij3}\sigma_i\sigma_j P_3^2. \tag{S.1d}$$

In accordance with LGD theory, the coefficient $\alpha$ depends linearly on the temperature $T$, $\alpha(T) = \alpha_T(T - T_C)$, where $T_C$ is the Curie temperature of the bulk ferrielectric. The coefficients $\beta, \gamma$, and $\delta$ in Eq.(S.1b) are temperature-independent. The values $g_{33kl}$ are polarization gradient coefficients in the matrix notation, and the subscripts $k, l = 1 - 3$. The values $\sigma_i$ denote diagonal components of a stress tensor in the Voigt notation, and the subscripts $i, j = 1 - 6$. The values $Q_{i3}$, $Z_{i33}$, and $W_{ij3}$ denote the components of a single linear and two nonlinear electrostriction strain tensors in the Voigt notation, respectively [44, 45]. $E_3$ is an electric field component co-directed with the polarization $P_3$, which is a superposition of external and depolarization fields.

**Table SI.** LGD parameters for a bulk ferroelectric CuInP$_2$S$_6$ at fixed stress

| Coefficient | Value |
|---|---|
| $\varepsilon_b$ | 9 |
| $\alpha_T$ (C$^{-2}$·m J/K) | 1.64067×10$^7$ |
| $T_C$ (K) | 292.67 |
| $\beta$ (C$^{-4}$·m$^5$J) | 3.148×10$^{12}$ |
| $\gamma$ (C$^{-6}$·m$^9$J) | −1.0776×10$^{16}$ |
| $\delta$ (C$^{-8}$·m$^{13}$J) | 7.6318×10$^{18}$ |
| $Q_{i3}$ (C$^{-2}$·m$^4$) | $Q_{13} = 1.70136 - 0.00363\,T$, $Q_{23} = 1.13424 - 0.00242\,T$, $Q_{33} = -5.622 + 0.0105\,T$ |
| $W_{ij3}$ (C$^{-2}$·m$^4$ Pa$^{-1}$) | $W_{113} \approx W_{223} \approx W_{333} \cong -2\times 10^{-12}$ |
| $Z_{i33}$ (C$^{-2}$·m$^4$) | $Z_{133} = -2059.65 + 0.8\,T$, $Z_{233} = -1211.26 + 0.45\,T$, $Z_{333} = 1381.37 - 12\,T$ |
| $s_{ij}$ (Pa$^{-1}$) | $s_{11} = 1.092\times 10^{-11}$, $s_{12} = -0.311\times 10^{-11}$, $s_{22} = 1.074\times 10^{-11}$ |
| $g_{33ij}$ (J m$^3$/C$^2$) | $g \cong 2\times 10^{-9}$ |

Using results of Kvasov and Tagantsev [46], one-component and one-dimensional approximations in Eq.(S.1), we made the Legendre transformation of Eq.(S.1) to the strain-polarization representation, $\tilde{G} = G + u\sigma$. Allowing for the equation of state (2), the renormalized free energy density is:

$$\tilde{g}_{LD+ES} = \frac{\alpha^*}{2}P_3^2 + \frac{\beta^*}{4}P_3^4 + \frac{\gamma^*}{6}P_3^6 + \frac{\delta^*}{8}P_3^8 - P_3 E_3. \tag{S.2}$$

The coefficients $\alpha^*, \beta^*, \gamma^*$, and $\delta^*$ in Eq.(S.2) depend on the mismatch strain, $u_m$, as:

$$\frac{\alpha^*}{2} = \frac{\alpha}{2} + \frac{u_m(Q_{23}(s_{12}-s_{11})+Q_{13}(s_{12}-s_{22}))}{s_{11}s_{22}-s_{12}^2} - u_m^2\frac{(s_{12}-s_{22})^2 W_{113}+(s_{11}-s_{12})^2 W_{223}}{2(s_{12}^2-s_{11}s_{22})^2}, \tag{S.3a}$$

$$\frac{\beta^*}{4} = \frac{\beta}{4} + \frac{Q_{23}^2 s_{11} - 2Q_{13}Q_{23}s_{12} + Q_{13}^2 s_{22}}{2(s_{11}s_{22}-s_{12}^2)} + u_m\left\{\frac{(s_{22}-s_{12})Z_{133}+(s_{11}-s_{12})Z_{233}}{s_{12}^2-s_{11}s_{22}} + \frac{Q_{23}(s_{12}(s_{12}-s_{22})W_{113}+s_{11}(s_{11}-s_{12})W_{223})+Q_{13}(s_{22}(-s_{12}+s_{22})W_{113}+s_{12}(-s_{11}+s_{12})W_{223})}{(s_{12}^2-s_{11}s_{22})^2}\right\}, \tag{S.3b}$$

$$\frac{\gamma^*}{6} = \frac{\gamma}{6} + \frac{(Q_{13}s_{22}-Q_{23}s_{12})Z_{133}+(Q_{23}s_{11}-Q_{13}s_{12})Z_{233}}{s_{11}s_{22}-s_{12}^2} -$$

$$\frac{-2Q_{13}Q_{23}s_{12}(s_{22}W_{113}+s_{11}W_{223})+Q_{23}^2(s_{12}^2 W_{113}+s_{11}^2 W_{223})+Q_{13}^2(s_{22}^2 W_{113}+s_{12}^2 W_{223})}{2(s_{12}^2-s_{11}s_{22})^2} +$$

$$u_m \frac{s_{22}^2 W_{113}Z_{133}+s_{11}^2 W_{223}Z_{233}-s_{12}(s_{22}W_{113}+s_{11}W_{223})(Z_{133}+Z_{233})+s_{12}^2(W_{223}Z_{133}+W_{113}Z_{233})}{(s_{12}^2-s_{11}s_{22})^2}, \qquad (S.3c)$$

$$\frac{\delta^*}{8} = \frac{\delta}{8} + \frac{s_{22}Z_{133}^2-2s_{12}Z_{133}Z_{233}+s_{11}Z_{233}^2}{2(s_{11}s_{22}-s_{12}^2)} + \frac{Q_{23}\big(s_{12}(s_{22}W_{113}+s_{11}W_{223})Z_{133}-s_{12}^2 W_{113}Z_{233}-s_{11}^2 W_{223}Z_{233}\big)}{(s_{12}^2-s_{11}s_{22})^2} +$$

$$\frac{Q_{13}(-s_{22}^2 W_{113}Z_{133}+s_{12}s_{22}W_{113}Z_{233}+s_{12}W_{223}(-s_{12}Z_{133}+s_{11}Z_{233}))}{(s_{12}^2-s_{11}s_{22})^2}. \qquad (S.3d)$$

Since $W_{ijk}$ are small enough, we remain only linear terms in $W_{ijk}$ in Eqs.(S.3), and omit all terms proportional to higher powers of the parameter. The higher powers of $W_{ijk}$ lead to the 12-th powers of polarization in the renormalized free energy density (S.2), making the mathematical complexity of results similar to Ref.[47]. These cumbersome and very small higher renormalization terms were account for numerically.

**Appendix B. Loops characterization for 2-4-6-8 LGD potential**

Polarization dependence on the external field, $P(E)$, follows from minimization of the LGD free energy with coefficients renormalized by the mismatch strain, and has the form:

$$E = \alpha^* P + \beta^* P^3 + \gamma^* P^5 + \delta^* P^7. \qquad (S.4)$$

Equation (S.4) describes the dependence of electric field $E$ on polarization $P$. Note that for the stability reasons one should suppose that $\delta^* > 0$. The first and second derivatives of Eq.(S.4) are

$$\frac{\partial E}{\partial P} = \alpha^* + 3\beta^* P^2 + 5\gamma^* P^4 + 7\delta^* P^6, \qquad (S.5)$$

$$\frac{\partial^2 E}{\partial P^2} = 6\beta^* P + 20\gamma^* P^3 + 42\delta^* P^5. \qquad (S.6)$$

The polarization $P_c$ corresponding to the coercive field $E_c$ is determined from the condition $\partial E/\partial P|_{P=P_c} = 0$:

$$\alpha^* + 3\beta^* P_c^2 + 5\gamma^* P_c^4 + 7\delta^* P_c^6 = 0. \qquad (S.7)$$

The coercive field obeys an obvious relation:

$$E_c = \alpha^* P_c + \beta^* P_c^3 + \gamma^* P_c^5 + \delta^* P_c^7 \qquad (S.8)$$

It is seen that Eq.(S.7) is bi-cubic, and its analytical solutions has a very cumbersome form. However, some critical points of hysteresis loops behavior with temperature, strain or other external parameters could be found, as summarized below:

I. Zero coercive field condition, $E_c = 0$, could be found from the system of equations:

$$\begin{cases} \alpha^* + 3\beta^* P_e^2 + 5\gamma^* P_e^4 + 7\delta^* P_e^6 = 0, \\ \alpha^* P_e + \beta^* P_e^3 + \gamma^* P_e^5 + \delta^* P_e^7 = 0. \end{cases} \qquad (S.9)$$

The nonzero solution for $P_e$ can be found from the first of Eq. (S.9) and has the form:

$$P_e = \pm\sqrt{-\frac{\gamma^*}{3\delta^*} + \frac{\sqrt{\gamma^{*2}-3\beta^*\delta^*}}{3\delta^*}}. \tag{S.10a}$$

$$P_e = \pm\sqrt{-\frac{\gamma^*}{3\delta^*} - \frac{\sqrt{\gamma^{*2}-3\beta^*\delta^*}}{3\delta^*}}. \tag{S.10b}$$

The second of Eqs.(S.9) for nonzero $P_e$ gives the following relation determining the point of zero coercive field:

$$\alpha^* + \beta^* P_e^2 + \gamma^* P_e^4 + \delta^* P_e^6 = 0 \tag{S.10c}$$

Note that the dashed curves in **Fig. 1** correspond to the condition of zero coercive field, i.e., they are calculated from Eqs.(S.10a) and (S.10c). The green solid curve in **Fig. 4(a)** is calculated from Eqs.(S.10a) and (S.10b). Dotted curves in **Fig. 1** correspond to the condition $\alpha^* = 0$.

II. The condition, when two values of the coercive field merge and disappear, corresponds to the point, where the first and the second derivatives of the polarization with respect to the electric field diverge simultaneously. The condition is

$$\begin{cases} \alpha^* + 3\beta^* P_d^2 + 5\gamma^* P_d^4 + 7\delta^* P_d^6 = 0, \\ 6\beta^* P_d + 20\gamma^* P_d^3 + 42\delta^* P_d^5 = 0. \end{cases} \tag{S.11}$$

The nonzero solution for $P_d$ could be found from the second of Eqs.(S.11):

$$P_d = \pm\sqrt{-\frac{5\gamma^*}{21\delta^*} + \frac{\sqrt{25\gamma^{*2}-63\beta^*\delta^*}}{21\delta^*}}. \tag{S.12a}$$

$$P_d = \pm\sqrt{-\frac{5\gamma^*}{21\delta^*} - \frac{\sqrt{25\gamma^{*2}-63\beta^*\delta^*}}{21\delta^*}}. \tag{S.12b}$$

The first equation (S.11) gives the condition of the loop disappearance:

$$\alpha^* + 3\beta^* P_d^2 + 5\gamma^* P_d^4 + 7\delta^* P_d^6 = 0. \tag{S.12c}$$

The dot-dot-dashed curves in **Fig. 1** are calculated from Eqs.(S.12a) and (S.12c), and the dot-dashed curves in **Fig. 1** are calculated from Eqs.(S.12b) and (S.12c).

## Appendix C. Description of the numerical algorithm for the search of hysteresis loops with negative slope

To ease the understanding of the algorithm explained in this part, the three types of loops we consider, as well as certain key points on them, are shown in the figure below.

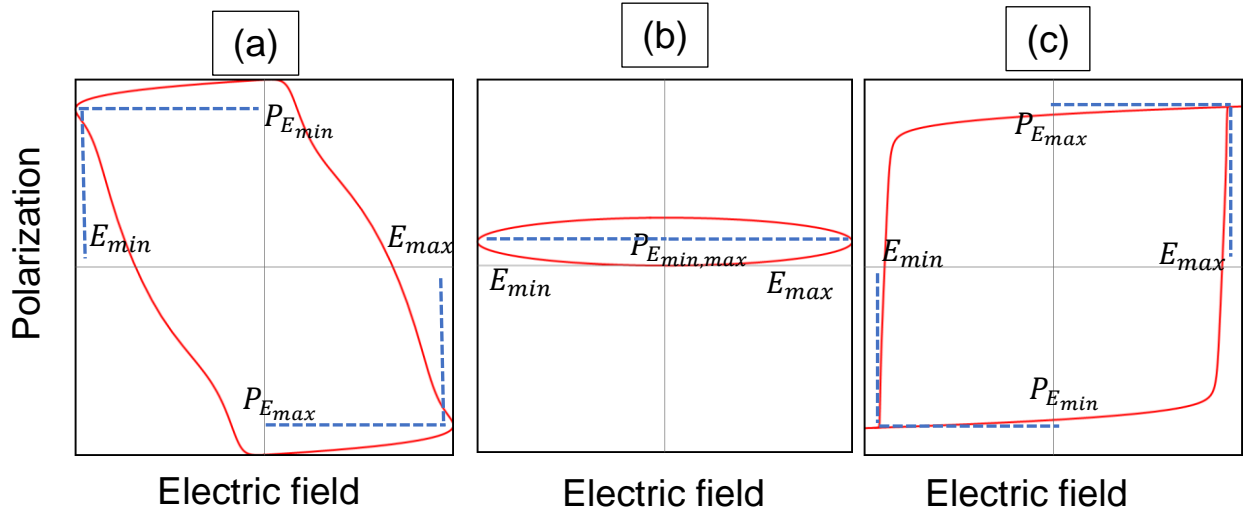

**FIGURE S1**. The three types of considered loops: a loop with a negative slope **(a)**, a partial loop **(b)**, and a "normal" loop with a positive slope **(c)**.

As one can see from the figure above, the following algorithm takes shape:

(a) for the loop with a negative slope, the polarization value at the minimal field $P_{E_{min}}$ is positive, and the polarization value at maximal field $P_{E_{max}}$ is negative:

$$P_{E_{min}} > 0 \ \& \ P_{E_{max}} < 0, \quad M = -1. \tag{S.13}$$

(b) For the loop with a positive slope, we have a completely opposite situation:

$$P_{E_{min}} < 0 \ \& \ P_{E_{max}} > 0, \quad M = 1. \tag{S.14}$$

(c) For a partial loop, both values are either positive or negative:

$$(P_{E_{min}} > 0 \ \& \ P_{E_{max}} > 0) \lor (P_{E_{min}} < 0 \ \& \ P_{E_{max}} < 0), \quad M = 0. \tag{S.15}$$

(d) For centrosymmetric elliptic, circular and quasi-circular loops, which should be excluded from our consideration, the condition is shown below:

$$\left|P_{E_{min}} - P_{E_{max}}\right| < \frac{\left|P_{E_{min}}\right|}{5}, \quad M = 2. \tag{S.16}$$

Next, for each case, a value of a model parameter $M$ is assigned if one of the conditions (S.13)-(S.16) is true. This parameter is evaluated for different frequencies and field amplitudes. Then, a contour map of this parameter is plotted and shown in **Figs. 6(c)-(e).**